\documentclass[11pt]{article}
\usepackage{amsthm, mathrsfs,amssymb,amsmath}
\usepackage{quantph} 
\newcommand{\abs}[1]{\vert #1 \vert}
\newcommand{\gen}[1]{\langle #1 \rangle}
\newcommand{\ket}[1]{\vert #1 \rangle}
\newcommand{\Integers}{\mathbb Z}
\newcommand{\ceil}[1]{\left\lceil #1 \right\rceil}
\newcommand{\floor}[1]{\left\lfloor #1 \right\rfloor}
\newtheorem{theorem}{Theorem}
\newtheorem{proposition}[theorem]{Proposition}
\newtheorem{definition}[theorem]{Definition}

\begin{document}

\begin{center} 
{\bf $\:$}\vspace{0mm}

{\LARGE \bf Quantum Property Testing of Group Solvability}\vspace{4mm}

{\Large Yoshifumi Inui$\:^{\star,\dagger}$ \hspace{15mm}Fran{\c c}ois Le Gall$\:^\dagger$}\vspace{6mm}

{\it $^\star$ Department of Computer Science, The University of Tokyo}\\
{\it 7-3-1 Hongo, Bunkyo-ku, Tokyo 113-0033, Japan} \vspace{1mm} 

{\it $^\dagger$ ERATO-SORST Quantum Computation and Information Project}\\
{\it Japan Science and Technology Agency}\\

{\it 5-28-3 Hongo, Bunkyo-ku, Tokyo 113-0033, Japan}\vspace{1mm}

email:  legall@qci.jst.go.jp\vspace{1mm}

\setlength{\baselineskip}{11pt}%
      \begin{quotation}
\noindent{\bf Abstract.}\hbox to .5\parindent{}
Testing efficiently whether a finite set $\Gamma$ with a binary operation $\cdot$ over it, 
given as an oracle, is a group is a well-known open problem in the field of property testing. 
Recently, Friedl, Ivanyos and Santha have made a significant step in the direction of solving this problem
by showing that it is possible to test efficiently whether the input $(\Gamma,\cdot)$ is an \emph{abelian} group 
or is far, with respect to some distance, from any abelian group. 
In this paper, we make a step further and construct an efficient quantum algorithm 
that tests whether $(\Gamma,\cdot)$ is a \emph{solvable} group, or is far from any solvable group.
More precisely, the number of queries used by our algorithm is polylogarithmic in the size 
of the set $\Gamma$.  
\end{quotation}
\end{center} 

\setcounter{footnote}{0}

\section{Introduction}
In property testing, the problem considered is to decide whether an object given as
an oracle has some expected property or is far from any object having that property.
This is a very active research area and many properties including algebraic function properties, 
graph properties, computational geometry properties and regular languages were proved to be testable.
We refer to, for example, \cite{kiwi,ron} for surveys on classical property testing.
Quantum testers have also been studied \cite{buhrman,fmss,magniez},
and they are known to be strictly more powerful than 
classical testers in some cases \cite{buhrman,magniez}.

In this paper, we focus on testing group-theoretical properties. 
A famous example is testing whether a function 
$f:G\to H$, where $H$ and $G$ are groups, is a homomorphism.
It is well known that such a test can be done efficiently \cite{ben-or,blr,shpilka}.
Another kind of problems deals with the case where the input is a finite set $\Gamma$ and
an oracle of a binary operation $\cdot:\Gamma\times\Gamma\to\Gamma$ over it. 
A classical algorithm testing associativity of the oracle $\cdot$ using $O(\abs{\Gamma}^2) $ 
queries to the oracle has been constructed 
by Rajagopalan and Schulman \cite{rajagopalan}, and
Erg\"un et al.~\cite{ergun} have proposed an algorithm, using $\tilde O(\abs{\Gamma})$ queries, testing if $\cdot$ is close to the multiplication of a group.
But notice that, since each element in $\Gamma$ needs $\mathrm\Theta(\log\abs\Gamma)$ bits 
to be encoded, the query complexities of these algorithms can be considered as exponential 
in the input length when not $\Gamma$, but only $\abs{\Gamma}$ is given 
(e.g.,~$\Gamma$ is supposed to be the set of binary strings of length $\ceil{\log_2\abs{\Gamma}}$).
Designing an algorithm deciding whether $(\Gamma,\cdot)$ is a group that uses a 
number of queries to $\cdot$ polynomial in 
$\log\abs{\Gamma}$ is indeed a well-known open problem.
Recently, Friedl et al.~\cite{friedl} have made a significant step in the direction of solving this problem by 
constructing a classical algorithm 
with query and time complexities polynomial in $\log\abs{\Gamma}$ 
that tests whether $(\Gamma,\cdot)$ is an abelian group or is far 
from any abelian group.

In this work, we make a step further and construct an efficient quantum algorithm 
that tests whether $(\Gamma,\cdot)$ is a solvable group or the distance between 
$(\Gamma,\cdot)$ and any solvable group is at least $\epsilon\abs{\Gamma}^2$.
More precisely, our algorithm uses a number of queries polynomial in 
$\log\abs{\Gamma}$ and $\epsilon^{-1}$, and its time complexity is polynomial in
$\exp((\log\log\abs\Gamma)^2)$ and $\epsilon^{-1}$, i.e.,~subexponential in $\log\abs\Gamma$. 
Notice that the class of solvable groups is far much
larger than the class of abelian groups and includes a vast class of non-abelian groups.
To deal with those groups, we introduce new ideas 
relying on the ability of quantum computation to solve fundamental
group-theoretical problems, such as finding orders of elements or
working with superpositions of all the elements of a subgroup. 

Besides the theoretical interest of this result, our algorithm 
can be used when studying group-theoretical problems 
where the input is a black-box solvable group (i.e.,~given
as a set a generators and an oracle performing group operations). 
Most known algorithms for such problems can have an unpredictable behavior when the input is not a solvable group. 
By applying our algorithm we can detect (in the quantum setting) if the input is far from any solvable group,
and we thus obtain robust versions of the 
quantum algorithms already known for solvable black-box groups \cite{Friedl+STOC03,inui,ivanyos,watrous}.
We also hope that this will be useful to 
design new quantum property testers or group-theoretical quantum algorithms.
In particular, our tester may be useful when considering quantum versions of 
classical algorithms solving problems over black-box solvable groups \cite{arvind,Babai+FOCS93,babai,babai2} as well.

Finally, we believe that our quantum algorithm may also be a first step in the direction of designing efficient classical
testers for solvable groups. Indeed, the efficient classical tester for abelian groups proposed by Friedl et 
al.~\cite{friedl} was inspired by a quantum algorithm solving the same problem. In this case, they were able 
to ``dequantumize'' the algorithm. A similar approach may be possible for our algorithm too.
   
\section{Definitions}\label{def}
\subsection{Distances between sets}\label{dist}
Let $\Gamma$ be a set and $\cdot: \Gamma\times \Gamma\to X$ a binary operation over it, where
$X$ is some set. We say that such couple $(\Gamma,\cdot)$ is a pseudo-magma.
If $X\subseteq\Gamma$, we say that $(\Gamma,\cdot)$ is a magma. 
When there is no ambiguity we will denote a pseudo-magma or a magma $(\Gamma,\cdot)$ simply by $\Gamma$.
We now define a distance between two 
pseudo-magmas. In this paper we adopt the so-called edit distance.
This is the same distance as the one used by Friedl et al.~\cite{friedl}.

Define a table of size $k$ as a $k\times k$ matrix with entries in some arbitrary set.
We consider three operations to transform a table to another.
An exchange operation replaces elements in a table by arbitrary elements and its cost is the number of replaced elements.
An insert operation at index $i$ inserts a row and a column of index $i$. Its cost is $2k+1$ if the original table
is of size $k$.
A delete operation at index $i$ deletes both the row of index $i$ and the column 
of index $i$, giving a table of size $(k-1)\times(k-1)$. Its cost is $(2k-1)$.

%
Let
$(\Gamma,\cdot)$ be a pseudo-magma, with $\cdot:\Gamma\times \Gamma\to X$. 
A multiplication table for $\Gamma$ is a table of size $\abs{\Gamma}$ with entries in $X$
for which both rows and columns are in one-to-one 
correspondence with elements in $\Gamma$, i.e.,~there exists a bijection 
$\sigma:\{1,\cdots,\abs \Gamma\}\rightarrow \Gamma$ 
such that the element in the 
$i$-th row and the $j$-th column is  $\sigma(i)\cdot\sigma(j)$.
The distance between two pseudo-magmas is defined as follows.
\begin{definition}
The edit distance between two tables $T$ and $T'$ is the minimum cost needed to transform $T$ to $T'$ by the above exchange, insert and delete operations. 
The edit distance between two pseudo-magmas $\Gamma$ and $\Gamma'$, 
denoted $d(\Gamma,\Gamma')$, is the minimum edit distance 
between $T$ and $T'$ where $T$ (resp.~$T'$) runs over all tables corresponding 
to a multiplication table of $\Gamma$ (resp.~$\Gamma'$).
For $\delta\ge 0$, we say that a pseudo-magma 
$\Gamma$ is $\delta$-close to another pseudo-magma $\Gamma'$ if 
$d(\Gamma,\Gamma')\le \delta$.
Otherwise we say that $\Gamma$ and $\Gamma'$ are $\delta$-far.
\end{definition}
Notice that if the sizes of $\Gamma$ and $\Gamma'$ are the same, then the edit distance becomes the minimal Hamming distance 
of the corresponding tables. 

\subsection{Property testing of group solvability}\label{propt}
In this paper we assume that the reader is familiar with the standard notions of group theory.
We refer to any standard textbook for details.
For completeness, we only recall the definition of solvable groups.
\begin{definition}
\label{def-sg}
A group $G$ is solvable if there exists a collection of subgroups $G_0, \ldots,G_k$ of $G$
such that:
\begin{itemize}
\item[(i)]
for each $0< j\le k$, the subgroup $G_{j-1}$ is normal in $G_j$ and $G_{j}/G_{j-1}$ is cyclic; 
\item[(ii)]
$\{e\}=G_0\lhd \cdots\lhd G_k=G$.
\end{itemize} 
\end{definition}

We now give our definition of a quantum property tester of group solvability. 
We define such a tester as a quantum algorithm $\mathscr{A}$ receiving as input a magma $(\Gamma,\cdot)$.
More precisely, the actual input of the algorithm is the value $\abs{\Gamma}$, and 
two oracles are available: an oracle that generates random elements in $\Gamma$ 
(the details of the implementation of this oracle are not essential because this oracle 
will only be used in a classical subprocedure), and a quantum oracle that performs the binary operation $\cdot$. 
Since the elements of $\Gamma$ can be encoded by binary strings of length 
$k=\lceil\log_2\abs{\Gamma}\rceil$, we identify the elements with their encoding and suppose that this 
quantum oracle performs the map $\ket{g}\ket{h}\ket{c}\mapsto \ket{g}\ket{h}\ket{c\oplus g\cdot h}$,
where $g$ and $h$ are elements in $\Gamma$ and $c$ is a string in $\{0,1\}^{k}$.
We denote by $\mathscr{A}(\Gamma)$ the behavior of the algorithm $\mathscr{A}$ on an input $(\Gamma,\cdot)$
given in this way. 
A more formal definition of a quantum property tester can be 
given but the following definition will be sufficient for our purpose.

\begin{definition}\label{definition_tester}
Let $d$ be the distance defined in Subsection \ref{dist}.
A quantum $\epsilon$-tester of group solvability is a quantum algorithm $\mathscr{A}$
such that, for any magma  $(\Gamma,\cdot)$, the following holds:
$$\left\{\begin{array}{l l l}
	{\bf Pr}[\mathscr{A}(\Gamma)\, accepts]>2/3\quad if\, d(\Gamma, \mathscr{S})=0\\
	{\bf Pr}[\mathscr{A}(\Gamma)\, rejects]>2/3\quad if\, d(\Gamma, \mathscr{S})>\epsilon \abs{\Gamma}^2.
\end{array}\right.$$
Here we use $d(\Gamma, \mathscr{S})$ to represent $\inf_{G\in \mathscr{S}}d(\Gamma,G)$, 
where $\mathscr{S}$ denotes the set of finite solvable groups.
\end{definition}

Notice that, a priori, requiring that the oracle is quantum may seem to give a problem different than in the classical setting,
where the oracle is classical. But this is not really the case: if a classical procedure that computes the 
product $g\cdot h$ from $g$ and $h$
is available, such a quantum oracle can be effectively constructed using standard techniques of quantum computation
\cite{Nielsen+00}. 

The main result of this paper is the following theorem.

\begin{theorem}\label{main_theorem}
There exists a quantum $\epsilon$-tester of group solvability that uses a number of queries polynomial in $\log \abs{\Gamma}$ and $\epsilon^{-1}$.
The running time of this algorithm is polynomial in $\exp((\log\log \abs{\Gamma})^2)$ and $\epsilon^{-1}$.
\end{theorem}

\subsection{Quantum algorithms for solvable groups}\label{section_group}

As stated in the following theorem,
efficient quantum algorithms for studying the structure of solvable groups
have been constructed by Watrous \cite{watrous}. Our algorithm deeply relies 
on these algorithms.

\begin{theorem}{(\cite{watrous})}\label{th_watrous}
Let $G$ be a solvable group given as a black-box group. 
Then there exists a quantum algorithm running
in time polynomial in $\log\abs{G}$ that outputs, with probability at least 3/4, 
$t=O(\log\abs{G})$ elements $h_1,\ldots,h_t$ of $G$ and $t$ integers $m_1,\ldots,m_t$ such that, if we denote $H_i=\gen{h_1,\ldots,h_i}$ for
$1\le i\le t$, the following holds. 
\begin{itemize}
\item[(a)]
$\{e\}=H_0\lhd H_1\lhd\cdots\lhd H_{t-1}\lhd H_t=G$; and
\item[(b)]
$H_{i}/H_{i-1}$ is cyclic, for $1\le i \le t$, with $\abs{H_i}/\abs{H_{i-1}}=m_i$.
\end{itemize}
Moreover, given any $0 \le i\le t$, and any element $g$ in $H_i$, there exists
a quantum algorithm running in time polynomial in $\log\abs{G}$ that outputs, with probability
at least $3/4$, 
the (unique) factorization of $g$ over $H_{i}$, i.e.,~integers
$a_1,\ldots,a_i$ with each $a_k\in\Integers_{m_k}$, such that $g=h_i^{a_i}h_{i-1}^{a_{i-1}}\cdots h_1^{a_1}$. 
\end{theorem}

In the algorithm of Theorem \ref{th_watrous}, the group is supposed to be input as a black-box group:
the input is a set of strings representing a set of generators of the group and 
an oracle performing the group product is available. The oracle necessary for Watrous's algorithm
\cite{watrous} 
is the map $\ket{g}\ket{h}\ket{c}\mapsto \ket{g}\ket{h}\ket{c\oplus g\cdot h}$,
for any elements $g,h\in G$ and any string $c$ in $\{0,1\}^{k}$. 
Notice that this is the same oracle as the one given to a quantum tester of group solvability as defined in
Subsection \ref{propt}. 

\section{Our Quantum Algorithm}
In this section we describe our quantum algorithm.
We first give an overview of the algorithm in Subsection \ref{subo}.
Then, in Subsection \ref{alg}, we explain the details.
Finally, we analyse its correctness and complexity
in Subsection \ref{anal}.
 
\subsection{Outline of our algorithm}\label{subo}
Our algorithm consists of four parts.\vspace{2mm}

\noindent {\bf Decomposition of $\Gamma$}\\
We first construct, using Theorem \ref{th_watrous},
$t=O(\log\abs\Gamma)$ elements $h_1,\ldots,h_t$ of $\Gamma$ that satisfy, if 
$\Gamma$ is a solvable group, the relations  
$
\{e\}=H_0\lhd H_1=\gen{h_1}\lhd\cdots\lhd H_i=\gen{h_1,\cdots,h_i}\lhd\cdots \lhd H_t=\gen{h_1,\cdots,h_t}=\Gamma,
$
where each $H_i$ is a subgroup of $\Gamma$, normal in $H_{i+1}$, such that $H_{i+1}/H_{i}$ is cyclic.
If $\Gamma$ is a solvable group, this decomposition gives a so-called power-conjugate presentation of $\Gamma$.
If $\Gamma$ is not a solvable group, these elements $h_1,\ldots,h_t$ will still define 
some pseudo-magmas $H_0,\ldots, H_t$, although in general these sets satisfy no group-theoretic property (in particular, they are not necessarily magmas).\vspace{2mm}

\noindent {\bf Test of embedding}\\
Then, we take sufficiently many elements of $\Gamma$ and check that they are all in $H_t$.
Success of this test implies that $\abs{\Gamma\backslash H_t}$ is small enough.
Of course, if $\Gamma$ is a solvable group, then $\Gamma=H_t$ with high probability 
and this test always succeeds.
Assume that $\Gamma$ is far from any solvable group $\tilde H_t$.
If the test succeed, since the inequality $d(\Gamma,\tilde H_t)\le d(\Gamma,H_t)+d(H_t, \tilde H_t)$ 
holds for any solvable group $\tilde H_t$, this will imply that
$H_t$ is far from any solvable group $\tilde H_t$ too 
(because the value of $d(\Gamma,H_t)$ is basically a function of $\abs{\Gamma\backslash H_t}$, and thus small).\vspace{2mm}

\noindent {\bf Construction of the group $G_t$}\\
We construct, using the information about the structure of $\Gamma$ obtained at the first part of the algorithm,
$t$ solvable groups $G_1,\ldots,G_t$ and a function $\psi:G_t\to H_t$
in a way such that, if $\Gamma$ is a solvable group, then $\psi$ is a group isomorphism from 
$G_t$ to $H_t$.\vspace{2mm}

\noindent {\bf Test of homomorphism}\\
Finally, the algorithm will test whether $\psi$ is ``almost'' an homomorphism.
We will show that this test is robust: if $\psi$ is close to an homomorphism, then
$H_t$ is close to the solvable group $G_t$. 
If $H_t$ is far from any solvable group, then this cannot hold and the homomorphism 
test must fail with high probability.\vspace{2mm}

Again, the similar idea of constructing a group $G$, a function $\psi:G\to \Gamma$ and 
use homomorphism tests was at the heart of the property tester for abelian groups 
proposed by Friedl et al.~\cite{friedl} and inspired this work 
(notice that the Friedl et al.~first constructed a quantum property tester for abelian groups,
and then were able to remove the quantum part in their algorithm). 
However there are new difficulties that arise when considering
property testers for solvable groups. The first one is that
analyzing the decomposition the $H_i$'s 
is more difficult and the power of quantum computation seems necessary to perform
this task efficiently.
The second complication is that, now, the groups $G_i$'s we are considering
are solvable, i.e.,~in general not commutative. In this case, we have to be 
very careful in the definition of $G_i$ and additional tests have to be done to ensure that 
the $G_i$'s we define are really groups. 

\subsection{Algorithm}\label{subsec_fig}
Our algorithm appears in Figure 1 and each of the four parts are explained 
in details in Subsections \ref{sub_prelim} to \ref{sub_homo}. 
If all the tests performed succeed, we decide that $\Gamma$ is a solvable group. 
Otherwise we decide that $\Gamma$ is $(\epsilon\abs{\Gamma}^2)$-far from any solvable group.

\begin{figure}[h]\label{alg}
\hrule\vspace{3mm}
{\bf PART I: Decomposition of $\Gamma$}\\
1.
Take $O(\log \abs\Gamma)$ random elements uniformly and independently in $\Gamma$.\\
2.
Use the first algorithm of Theorem \ref{th_watrous} on them and obtain the set $\{h_1,\ldots,h_t\}$
and integers $m_1,...,m_t$. \\
3. 
For each $i\in\{1,\ldots,t\}$, use Shor's order finding algorithm on  $h_i$ and obtain some integer $n_i$.\\
4.
Compute the decompositions of all $h_i^{m_i}$ and $h_i^{n_i-1}\cdot(h_k\cdot h_i)$ over $H_{i-1}$,
for $i\in\{1,\ldots,t\}$ \\
$\phantom{3. .}$ and $k\in\{1,\ldots,i-1\}$, and check the obtained decompositions.\vspace{2mm} 

\noindent{\bf PART II: Test of embedding}\\
5.
Check that $\abs{\Gamma}=m_1\times\cdots\times m_t$ and
$\abs{\Gamma\backslash H_t}/\abs{\Gamma}<\epsilon/4$.\vspace{2mm}

\noindent{\bf PART III: Construction of the group $G_t$}\\
6.
For $j$ from 2 to $t$ check that Conditions (a), (b) and (c) of Proposition \ref{proposition_threeconditions} hold. \vspace{2mm}

\noindent{\bf PART IV: Test of homomorphism}\\
7.
Check that ${\bf Pr}_{x, y\in G_t}[\psi(x\circ y)=\psi(x)\cdot \psi(y)]>1-\eta$ with $\eta=\epsilon/422$.\\
\hrule
\caption{Quantum $\epsilon$-tester of group solvability}
\end{figure}


\subsubsection{Decomposition of $\Gamma$}\label{sub_prelim}
$\phantom{s}$\newline
The first step in our algorithm finds a power-conjugate representation of 
$\Gamma$ when $\Gamma$ is a solvable group. 
We will prove that when $\Gamma$ is far from any solvable group, 
then the output of this step cannot be a power-conjugate 
representation of a group close to $\Gamma$ and that this can be detected by our algorithm
at part II, III or IV.

We begin by picking $s=\mathrm\Theta(\log \abs\Gamma)$ random elements $\alpha_1, \cdots, \alpha_s$ uniformly and independently from the ground set $\Gamma$. 
For simplicity, we first suppose  that $\Gamma$ is a solvable group, and then discuss the general case.

\paragraph{Case where $\Gamma$ is a solvable group.}
Denote $\Gamma'=\gen{\alpha_1, \cdots, \alpha_s}$.
Then, with high probability, 
$\Gamma=\Gamma'$. 
Here we rely on the standard fact in computational group theory that, for any group $K$, $\mathrm\Theta(\log\abs K)$ random elements taken uniformly 
in $K$ constitute, with high probability, a generating set of $K$.
We now run the first algorithm of Theorem \ref{th_watrous} with input
$\Gamma'$ presented as a black-box group as follows: $\alpha_1, \cdots, \alpha_s$ is the set of generators and
the operation $\cdot$ is the oracle performing group multiplication. The output of the algorithm is then, with
high probability, a set of $t$ elements $h_1,\ldots,h_t$ of $\Gamma$ and $t$ integers $m_1,\ldots,m_t$ such that, 
if we denote $H_i=\gen{h_1,\ldots,h_i}$ for
$1\le i\le t$, the following holds:
\begin{itemize}
\item[(a)]
$\{e\}=H_0\lhd H_1\lhd\cdots\lhd H_{t-1}\lhd H_t=\Gamma'$; and
\item[(b)]
$H_{i}/H_{i-1}$ is cyclic for $1\le i \le t$ and satisfies $\abs{H_i}/\abs{H_{i-1}}=m_i$.
\end{itemize}
We then use Shor's quantum algorithm \cite{Shor} to compute the order $n_i$ of each $h_i$ in $\Gamma$.
Moreover, we further analyze the structure of $\Gamma'$ and use the second algorithm
of Theorem \ref{th_watrous} to decompose the elements 
$h_i^{m_i}$ and  $h_i^{n_i-1}\cdot(h_k\cdot h_i)$ over $H_{i-1}$, 
for each $i\in\{2,\ldots,t\}$ and each $k\in\{1,\ldots,i-1\}$. 
Notice that, indeed,  each $h_i^{m_i}$ and 
$h_i^{n_i-1}\cdot(h_k\cdot h_i)=h_i^{-1}\cdot h_k\cdot h_i$ are in $H_{i-1}$
when $\Gamma$ is a solvable group.
We denote the decompositions obtained by 
\begin{equation}\label{rel1}
h_i^{m_i}=h_{i-1}^{r^{(i)}_{i-1}}\cdot\left(
\cdots \cdot\left(h_{3}^{r^{(i)}_{3}}\cdot\left(h_{2}^{r^{(i)}_{2}}\cdot h_{1}^{r^{(i)}_{1}}\right)\right)\right)
\textrm{ for }2\le i \le t,
\end{equation}
\begin{equation}\label{rel2}
h_{i}^{n_i-1}\cdot(h_k\cdot h_i)=
h_{i-1}^{s^{(i)}_{k,i-1}}\cdot\left(
\cdots \cdot\left(h_{3}^{s^{(i)}_{k,3}}\cdot\left(h_{2}^{s^{(i)}_{k,2}}\cdot h_{1}^{s^{(i)}_{k,1}}\right)\right)\right)
\textrm{ for }1\le k<i\le t,
\end{equation}
where each $r_{\ell}^{(i)}$ and each $s_{k,\ell}^{(i)}$ are in $\Integers_{m_\ell}$.
(The parentheses are superfluous when $\cdot$ is associative, but not in the general case we discuss below.)
 
 \paragraph{General Case.}
In general, we do not know whether $\Gamma$ is a solvable group or not but we do exactly the same as above:
we first run the first algorithm of Theorem \ref{th_watrous} on the set $\{\alpha_1, \cdots, \alpha_s\}$ with the oracle $\cdot$. If
this algorithm errs, 
we conclude
that $\Gamma$ is not a solvable group
(this decision is correct with high probability because, if 
$\Gamma$ is a solvable group, then the algorithm of Theorem \ref{th_watrous} succeeds with high probability). 
Now suppose that we have obtained elements $h_1,\ldots,h_t$ and a set of
integers $m_1,\ldots,m_t$. 
We define the following sets by recurrence:
$H_1=\{h_1^a\vert a\in\mathbb Z_{m_1}\}$,  and, for $2\le j\le t$, 
$H_j=\{h_j^a\cdot h\vert a\in\mathbb Z_{m_j}, h\in H_{j-1}\}$. 
Here, and in many other places in this paper, 
we use the notation $h^r$, for $h\in\Gamma$ and $r\ge 1$, to denote the product $h\cdot(\cdot\cdots 
(h\cdot(h\cdot h)))$, since $\cdot$ is not in general associative. 
Moreover we use the convention $h^0=h_1^{m_1}$ for any $h\in\Gamma$.
Notice that the value of $h^r$ can be computed using $O(\log r)$ queries to the oracle $\cdot$ using repeated squaring methods.

Notice that, in general, the pseudo-magmas $H_i$'s have no group-theoretical structure at all (in particular they may not be magmas).
We then use Shor's order finding algorithm \cite{Shor} on each $h_i$ and obtain some integer $n_i$.
Then we run the second algorithm of Theorem \ref{th_watrous} to decompose the elements 
$h_i^{m_i}$ and $h_i^{n_i-1}\cdot(h_k\cdot h_i)$ over $H_{i-1}$, 
for each $i\in\{2,\ldots,t\}$ and each $k\in\{1,\ldots,i-1\}$.
If the algorithm errs or outputs something irrelevant, we conclude that $\Gamma$ is not a solvable group.
Suppose that the algorithm succeeds and outputs decompositions. We use the notations of Equations (\ref{rel1}) and (\ref{rel2})
to denote the decompositions obtained. We check whether these decompositions are correct, i.e.,~we compute 
the right sides of Equations (\ref{rel1}) and (\ref{rel2}) and check that they match the left sides.
 If they are correct, we move to the next step (Subsection \ref{sub_emb}). Otherwise, we conclude that 
$\Gamma$ is not a solvable group.
 
\subsubsection{Test of embedding}\label{sub_emb}
$\phantom{s}$\newline
In the second part of our algorithm, we first check that $\abs{\Gamma}=m_1\times\cdots\times m_t$.
Then, we want to check whether $\abs{\Gamma\backslash H_t}$ is small
enough. Otherwise we conclude that $\Gamma$ is not a solvable group.
Indeed, if $\Gamma$ is a group, then with high probability (on the choice of $\alpha_1,\ldots,\alpha_s$
and on the randomness of the algorithm of Theorem \ref{th_watrous}) $\Gamma=H_t$.

More precisely we check whether 
$\abs{\Gamma\backslash H_t}/\abs{\Gamma}<\epsilon/4$ holds.
In order to perform this test, we simply take $c_1$ elements of $\Gamma$ and check whether 
they are all in $H_t$ (by using the second algorithm of Theorem \ref{th_watrous} and checking 
the obtained decompositions). It is easy to show that, when taking $c_1=\mathrm\Theta(\epsilon^{-1})$, 
we can detect whether $\abs{\Gamma\backslash H_t}/\abs{\Gamma}>\epsilon/4$ 
with constant probability.

\subsubsection{Construction of the group $G_t$}\label{const}
$\phantom{s}$\newline
We now show how to construct an abstract group $G_t$ defined by 
the power-conjugate presentation found in Part I of our algorithm
(Equations (\ref{rel1}) and (\ref{rel2})) when such a group
exists, i.e.,~when the presentation is consistent with the definition
of a group.

We first define by recurrence the family of magmas $\{G_j\}_{1\le j\le t}$, where each $G_j$ is equal 
(as a set) to $\Integers_{m_{j}}\times\cdots\times \Integers_{m_{1}}$. 
$G_1$ is defined as the cyclic group $(\Integers_{m_1},+)$, where
$+$ is the addition modulo $m_1$.
For any $i\in\{2,\ldots,t\}$, denote by $u_i$ the element $(r^{(i)}_{i-1},\ldots,r^{(i)}_1)$ of $G_{i-1}$ and,
for any $i\in\{2,\ldots,t\}$ and $k\in\{1,\ldots,i-1\}$, denote by $v_{i,k}$ 
the element $(s^{(i)}_{k,i-1},\ldots,s^{(i)}_{k,1})$ of $G_{i-1}$.

\begin{definition}\label{def_G}
Define $G_1=(\Integers_{m_1},+)$ and, for $2\le j\le t$, let $G_j$ be the magma 
$(\mathbb Z_{m_j}\times G_{j-1},\circ_j)$ with 
$$(a, x)\circ_j(b, y)=
\left\{
\begin{array}{ll}
\left(a+b, \phi_{j}^{(b)}({x})\circ_{j\!-\!1}y\right)&\textrm{ if }\:\: a+b<m_j\\
\left(a+b-m_j, u_j\circ_{j\!-\!1}\phi_{j}^{(b)}({x})\circ_{j\!-\!1}y\right)&\textrm{ if }\:\: a+b\ge m_j
\end{array}\right.
$$
where $\phi_{j}:G_{j-1}\to G_{j-1}$ maps any element $(a_{j-1},\cdots,a_1)$ of $G_{j-1}$ to the element  
$\phi_j((a_{j-1},\cdots,a_1))=v_{j,j-1}^{a_{j-1}}\circ_{j\!-\! 1} \left(\cdots 
\circ_{j\!-\! 1} \left(v_{j,2}^{a_{2}}\circ_{j\!-\!1}v_{j,1}^{a_{1}}\right)\right)$ of $G_{j-1}$, and $\phi_j^{(b)}$
means $\phi_j$ composed by itself $b$ times.
\end{definition}
We will usually denote $\circ_j$ or $\circ_{j\!-\! 1}$ simply by $\circ$ when there is no ambiguity.

In order to illustrate this definition, let us consider the case where all the $H_j$'s are solvable groups.
In this case, each $H_j=\{h_j^{a_j}\cdot\cdots \cdot h_1^{a_1}\:|\:a_j\in\Integers_{m_j}\}$ is in bijection with 
$\Integers_{m_{j}}\times\cdots\times \Integers_{m_{1}}$ (as a set).
Fix a $j$ and consider $H_j$. Each element $h_j^{a_j}\cdots h_1^{a_1}$ is associated with the element 
$(a_j,\ldots,a_1)$ of $G_j$. Now the element $\phi_{j}((a_{j-1},\cdots,a_1))$ corresponds to the element
$$h_j^{-1}\cdot (h_{j-1}^{a_{j-1}}\cdots h_{1}^{a_1})\cdot h_j=
\left(h_{j-1}^{s^{(j)}_{j-1,j-1}}\cdots h_{1}^{s^{(j)}_{j-1,1}}\right)^{a_{j-1}}\cdots 
\left(h_{j-1}^{s^{(j)}_{1,j-1}}\cdots h_{1}^{s^{(j)}_{1,1}}\right)^{a_{1}}.
$$
In other words, the map $\phi_{j}$ in $G_{j-1}$ corresponds to the 
automorphism $h\mapsto h_j^{-1}h h_j$ of $H_j$.  
For any two elements $g$ and $g'$ in $H_{j-1}$, since 
$h_j^a\cdot g\cdot h_j^b\cdot g'=h_j^{a+b}\cdot(h_j^{-b}\cdot g\cdot h_j^b)\cdot g'$ we see that
the $G_j$'s are defined to be isomorphic to the $H_j$'s in the case where the $H_j$'s are solvable groups.

If the $H_j$'s are not groups, then the $G_j$'s constructed in Definition \ref{def_G}
are not necessarily groups. But we now show that when 
some additional conditions are satisfied, the $G_j$'s become groups. In technical words these are 
necessary and sufficient conditions to make the presentation of $G_j$ a consistent presentation of 
successive cyclic extensions. 
In the next proposition, we denote by $x_{j,k}$, for $1\le k\le j\le t$, 
the element of $G_{j}$ with one $1$ at the index $k$ (from the right) and zeros at all the other indexes.

\begin{proposition}\label{proposition_threeconditions}
Let $1<j<t$.
Suppose that $G_{j-1}$ is a solvable group and, if $j\ge 3$, suppose additionally that $G_{j-2}$ is a solvable group and $\phi_{j-1}$ is a group automorphism of $G_{j-2}$. 
Assume that
the following three conditions hold.
\begin{itemize}
\item[(a)]
$x_{j-1,k}\circ v_{j-1,j-1}=
v_{j-1,j-1}\circ v_{j-1,k}$
for all $1\le k<  j-1$; and
\item[(b)]
 $\phi_j(u_j)=u_j$; and 
\item[(c)]
$\phi^{(m_j)}_j(x_{j-1,i})=u^{-1}_j\circ x_{j-1,i}\circ u_j$
for all $1\le i\le j-1$.
\end{itemize}
Then $G_j$ is a solvable group and $\phi_{j}$ is a group automorphism of $G_{j-1}$.
\end{proposition}
\begin{proof}
If $\phi_j$ is an automorphism of $G_{j-1}$, then Conditions (b) and (c)
imply that $G_j$, as defined in Definition \ref{def_G}, is a so-called cyclic extension of $G_{j-1}$ and thus a solvable group
(see for example \cite[Section 9.8]{Sims94}). 
We will show below that Condition (a) implies that $\phi_j$ is an endomorphism of $G_{j-1}$.
Since $\phi_j^{(m_j)}$ is an automorphism of $G_{j-1}$ from Condition (c), $\phi_j$ is thus
an automorphism too.

We now prove that $\phi_j$ is an endomorphism of $G_{j-1}$. 
If $j=2$, then this is obviously the case: $\phi_2$ is the endomorphism of $G_1=(\Integers_{m_1},+)$ mapping $a$ to $av_{11}^{(2)}$.
In the following we suppose that $j\ge 3$.
We first start with a few useful observations.
First notice that, for any $a$ and $b$ in $\Integers_{m_{j-1}}$, the equality $\phi_j((a+b,e))=\phi_j((a,e))\circ\phi_j((b,e))$, where $e$ denotes
the unity element of $G_{j-2}$, holds from the definition of $\phi_j$.
Also notice that, for any $a$ in $\Integers_{m_{j-1}}$ and any $x$ in $G_{j-2}$, the equality $\phi_j((a,x))=\phi_j((a,e))\circ\phi_{j-1}(x)$ holds.

Any element $z\in G_{j-2}$ can be written in the form
$z=x_{j-1,j-2}^{\alpha_{j-2}}\cdots x_{j-1,1}^{\alpha_{1}}$ for some integers $\alpha_{1},\ldots,\alpha_{j-2}$.
Condition (a) then implies that the equality
$$
z\circ v_{j-1,j-1}=v_{j-1,j-1}\circ 
v_{j-1,j-2}^{\alpha_{j-2}}\circ\cdots \circ
v_{j-1,1}^{\alpha_{1}}=v_{j-1,j-1}\circ\phi_{j-1}(z)
$$
holds (since $\phi_{j-1}$ is an endomorphism of $G_{j-2}$ and $\phi_{j-1}(x_{j-1,k})=v_{j-1,k}$ for any $1\le k< j-1$).
More generally, for any $b\in\Integers_{m_{j-1}}$ and any $z\in G_{j-2}$, we have
$$
z \circ \phi_j((b,e))=z\circ v_{j-1,j-1}^b=v_{j-1,j-1}^b\circ \phi_{j-1}^{(b)}(z)=\phi_j((b,e))\circ \phi_{j-1}^{(b)}(z).
$$

Let $a, b$ be two elements of $\Integers_{m_{j-1}}$ and $x,y$ be two elements of $G_{j-2}$.
Putting together the above observations we can write 
\begin{eqnarray*}
\phi_j((a,x))\circ\phi_j((b,y))&=&\phi_j((a,e))\circ\phi_{j-1}(x)\circ\phi_j((b,e))\circ\phi_{j-1}(y)\\
&=&\phi_j((a,e))\circ\phi_j((b,e))\circ\phi_{j-1}^{(b+1)}(x)\circ\phi_{j-1}(y)\\
&=&\phi_j((a,e))\circ\phi_j((b,e))\circ\phi_{j-1}(\phi_{j-1}^{(b)}(x)\circ  y)\\
&=&
\phi_j((a,e))\circ\phi_j((b,\phi_{j-1}^{(b)}(x)\circ y))\\
&=&\phi_j((a+b,v\circ \phi_{j-1}^{(b)}(x)\circ  y)),
\end{eqnarray*}
where $v=u_j$ if $a+b\ge m_j$ and $v=e$ otherwise. We conclude that 
$$
\phi_j((a,x))\circ\phi_j((b,y))=
 \phi_j((a,x)\circ (b,y)),$$
and thus $\phi_j$ is an endomorphism of $G_{j-1}$.
\end{proof}

To illustrate the three conditions of Proposition \ref{proposition_threeconditions},
let us again consider the case where $(\Gamma,\cdot)$ is a group. Then conditions 
(b) and (c) hold due to the facts that $u_j$ in $G_{j-1}$ corresponds to the element $h_j^{m_j}$  
and that $\phi_{j}$  corresponds to the  automorphism $h\mapsto h_j^{-1}h h_j$ of $H_{j-1}$. 
Condition (a) follows from Equation (\ref{rel2}).

For each $j\in\{2,\ldots,t\}$, testing that Conditions (a) and (b) hold can be done 
using a number of multiplications in the group $G_{j-1}$ polynomial in $\log\abs{\Gamma}$.
The best known classical algorithm for computing products in a solvable group
given as a power-conjugate presentation
is an algorithm by H{\"o}fling \cite{hoefling} with time complexity 
$O(\exp((\log\log\abs{G_{j-1}})^2))=O(\exp((\log\log\abs\Gamma)^2))$.
Notice that if Condition (a) holds then $\phi_j$ is a homomorphism.
Then each term $\phi^{(m_j)}_j(x_{j-1,i})$ in Condition (c) can be computed
using a number of group products polynomial in $\log\abs{\Gamma}$
by computing, step by step by increasing $\ell$ from $0$ to $\floor{\log m_j}$,
the values $\phi_{j}^{(2^\ell)}(x_{j-1,k})$ for all $1\le k\le j-1$.
The total time complexity of checking that all the $G_i$'s are solvable 
groups is 
thus $O(\exp((\log\log\abs\Gamma)^2))$. No query to the oracle $\cdot$ 
is needed.

\subsubsection{Test of homomorphism}\label{sub_homo}
$\phantom{s}$\newline
We now suppose that the $G_i$'s have passed all the tests 
of Proposition \ref{proposition_threeconditions} and thus 
$G_t$ is a solvable group.
Let $\psi$ be the surjective map from $G_t$ to  $H_t$ defined as
$$\psi(a_t,a_{t-1},\cdots,a_1)=h_t^{a_t}\cdot (h_{t-1}^{a_{t-1}}
\cdot(\cdots \cdot (h_{2}^{a_2}\cdot h_{1}^{a_1})).$$
We will test whether
$\psi$ is a homomorphism from $G_t$ to $H_t$.
If $(\Gamma,\cdot)$ is a solvable group, then $\psi$ is an homomorphism by construction.
We now show that this test is robust.
\begin{proposition}\label{proposition_closeness}
Let $\eta$ be a constant such that $0<\eta<1/120$. 
Assume that $\abs{H_{t}}> 3\abs{G_t}/4$. 
Suppose that 
\begin{equation}\label{condition_homo}
{\bf Pr}_{x, y\in G_t}[\psi(x\circ y)=\psi(x)\cdot \psi(y)]>1-\eta .
\end{equation}
Then there exists a solvable group $\tilde H_t$ that is $(211\eta\abs{\Gamma}^2)$-close to $H_t$.
\end{proposition}
\begin{proof}
From Condition (\ref{condition_homo}), 
Theorem 2 of \cite{friedl} implies that 
there exists a group $(\tilde H_t,*)$ with $\abs{\tilde H_t}\le \abs{G_t}$, and a homomorphism 
$\tilde \psi:G_t\rightarrow\tilde H_t$  such that:
\begin{enumerate}
\item[(a)] $\vert\tilde H_t\backslash H_t\vert\le30\eta\abs{\tilde H_t}$;
\item[(b)] ${\bf Pr}_{h, h'\in\tilde H_t}[h*h'\ne h\cdot h']\le91\eta$; and
\item[(c)] ${\bf Pr}_{x\in G_t}[\tilde \psi(x)\ne \psi(x)]\le30\eta$.
\end{enumerate}
Notice that, strictly speaking, Theorem 2 of \cite{friedl} 
is stated only in the case where $H_t$ is a magma, i.e.,~closed under $\cdot$. 
This is not the case here because $H_t$ may not be a magma, but only a pseudo-magma. 
However, careful inspection of the proof of Theorem 2 of \cite{friedl} shows that exactly the same result 
holds when $H_t$ is a pseudo-magma too.
The distance between $\tilde H_t$ and $H_t$ is determined by the number of elements being a member of either set and
the number of pairs of two elements for which the result of the multiplication differ.
In particular, this distance has for upper bound the cost of the following transform: starting from the table of $\tilde H_{j}$,
we first delete rows and columns corresponding to elements in $\tilde H_t\backslash H_t$,
insert rows and columns corresponding to elements in $H_t\backslash\tilde H_t$,
and then exchange multiplication entries which differ between two tables.
It follows from (a) and (b)  that the number of elements in $\tilde H_t\backslash H_t$ is
less than $30\eta\abs{\tilde H_t}$ and the number of pairs
$(h,h')\in\tilde H_t\times\tilde H_t$ such that $h* h'\ne h\cdot h'$ is 
less than $91\eta\abs{\tilde H_t}^2$.
It remains to show that $H_t\backslash \tilde H_t$ is small enough too and
that $\tilde H_t$ is a solvable group.

Suppose towards a contradiction that $\abs{\tilde \psi(G_t)}<\abs{G_t}$. Then
$\abs{\tilde\psi(G_t)}\le \abs{G_t}/2$. From Condition (c), we obtain 
$\abs{H_t}=\abs{\psi(G_t)}\le \abs{G_t}/2+30\eta\abs{G_t}\le 3\abs{G_t}/4$. This gives a 
contradiction. Thus $\abs{\tilde \psi(G_t)}=\abs{\tilde H_t}=\abs{G_t}$ and $\tilde\psi$ is an 
isomorphism from $G_t$ to $\tilde H_t$. Since $G_t$ is a solvable group, $\tilde H_t$
is solvable too. Since $\abs{H_{t}}\le \abs{G_t}$, 
it also follows that $\abs{H_t}\le\abs{\tilde H_t}$ and thus 
$\abs{H_t\backslash\tilde H_{t}}\le\abs{\tilde H_t\backslash H_{t}}\le 30\eta \abs{\tilde H_t}$.

Deleting $\abs{\tilde H_t\backslash H_t}$ rows and column from the table of $\tilde H_t$ costs
$$2\abs{\tilde H_t}\abs{\tilde H_t\backslash H_t}-\abs{\tilde H_t\backslash H_t}^2\le 60\eta\abs{\tilde{H_t}}^2.$$
Then inserting $\abs{H_t\backslash \tilde H_t}$ rows and columns similarly costs at most 
$60\eta\abs{\tilde{H_t}}^2$ too. 
Thus the distance between $H_t$ and the solvable group $\tilde H_t$ is at most 
$[(60+60+91)\eta\abs{\tilde H_t}^2]\le 211\eta\abs{\Gamma}^2.$
\end{proof}

More precisely, we perform the following test.
We want to test which of ${\bf Pr}_{x, y\in G}[\psi(x\circ y)=\psi(x)\cdot \psi(y)]=1$
and ${\bf Pr}_{x, y\in G_t}[\psi(x\circ y)=\psi(x)\cdot \psi(y)]\le 1-\eta$ with $\eta=\epsilon/422$ holds.
We  take $c_2$ pairs $(x,y)$ of elements of $G_t$ and test whether 
they all satisfy $\psi(x\circ y)=\psi(x)\cdot \psi(y)$.
It is easy to show that, when taking $c_2=\mathrm\Theta(\eta^{-1})=\mathrm\Theta(\epsilon^{-1})$, 
we can decide which case holds 
with constant probability.

\subsection{Correctness and complexity}\label{anal}
We now evaluate the performance of our algorithm.
This gives the result of Theorem \ref{main_theorem}.

First, suppose that the magma $(\Gamma,\cdot)$ is a solvable group.
With high probability the set of elements taken at step 1 of the algorithm of Figure 1 is a generating set of $\Gamma$ 
and the first algorithm of Theorem \ref{th_watrous} succeeds on this set. In this case, 
each of the tests realized at steps 3 to 5 succeeds with high probability
(since the success probability of Shor's algorithm and of the second algorithm of Theorem \ref{th_watrous} can be 
amplified), and then all the tests at steps 6 and 7 succeed with probability 1.
Thus the global error probability is constant.

Now, we would like to show that any magma $\Gamma$ that is 
$(\epsilon\abs{\Gamma}^2)$-far from any solvable group 
is rejected with high probability.
Take such a magma $\Gamma$.
Then $H_t$ is $(\frac\epsilon 2\abs{\Gamma}^2)$-far from any solvable group $\tilde H_t$ or 
$\abs{\Gamma\backslash H_t}/\abs\Gamma>\epsilon /4$.
This assertion holds because for any solvable group $\tilde H_t$, the inequalities
$\epsilon\abs{\Gamma}^2<d(\Gamma,\tilde H_t)\le d(\Gamma,H_t)+d(H_t, \tilde H_t)$ hold
and 
$d(\Gamma, H_t)=2\abs{\Gamma\backslash H_t}\abs{\Gamma}-\abs{\Gamma\backslash H_t}^2
\le 2\abs{\Gamma\backslash H_t}\abs{\Gamma}$ since $H_t\subseteq \Gamma$ and the operation is the same. 
If the latter holds, it should be rejected with high probability at test 5.
Now suppose that the former holds and that all the steps 1--6 succeed.
Then with high probability $\abs{H_t}\ge (1-\epsilon/4)\abs{\Gamma}\ge 3\abs{\Gamma}/4=3\abs{G_t}/4$. 
From Proposition \ref{proposition_closeness} this implies that 
${\bf Pr}_{x, y\in G_t}[\psi(x\circ y)=\psi(x)\cdot \psi(y)]\le 1-\epsilon/422$.
This is detected with high probability at step 7.

The algorithm queries the oracle $\Gamma$ a number of 
times polynomial in $\log \abs \Gamma$ at each of the steps 1 to 4, and
a number of times polynomial in $\log \abs{\Gamma}$ and $\epsilon^{-1}$
at steps 5 
and 7. 
Additional computational work is needed at steps 6 and 7 to compute a polynomial number 
of products in the groups $G_i$'s. Since each product can be done (without queries) using 
$O(\exp((\log\log\abs{G_i})^2))=O(\exp((\log\log \abs\Gamma)^2))$ time using the algorithm 
by H{\"o}fling \cite{hoefling}, the total time complexity of the algorithm is polynomial 
in $\exp((\log\log \abs \Gamma)^2)$ and $\epsilon^{-1}$.
 
\section*{Acknowledgments}
The authors thank anonymous reviewers for helpful comments and suggestions.

\end{document}